\documentclass[showpacs,aps, twocolumn,nofootinbib]{revtex4-2}
\usepackage{epsfig}
\usepackage{graphicx}
\usepackage{amsmath,amssymb,amsfonts}
\usepackage{array}
\usepackage{url}
\usepackage{xcolor}
\usepackage[colorlinks=true,linkcolor=purple, citecolor = blue]{hyperref}
\usepackage{multirow}
\usepackage{braket}
\usepackage{csquotes}
\usepackage{float}
\usepackage{lineno}
\usepackage{xspace}
\usepackage{ulem}
\usepackage{lipsum} 
\usepackage{footnote}
\begin{document}

\title{Rotational Brownian Motion and $\mathrm{J/\psi}$ Spin Alignment in Heavy-Ion Collisions}

\author{Bhagyarathi Sahoo$^{1}$}
\email{Bhagyarathi.Sahoo@cern.ch}
\author{Captain R. Singh$^{2}$}
\email{captainriturajsingh@gmail.com}
%
%
\affiliation{$^1$Department of Physics, Indian Institute of Technology Indore, Simrol, Indore 453552, India}
\affiliation{$^2$School of Physical Sciences, National Institute of Science Education and Research,
Jatni, Odisha–752050, India}

\begin{abstract}
The heavy-quark polarization in ultra-relativistic heavy-ion collisions (HICs) serves as a probe for unfolding the characteristics of deconfined QCD. The present study explores the spin alignment of $\mathrm{J/\psi}$ in the HICs by employing the rotational Brownian motion of charm quarks in the presence of a strong magnetic field. We derive an analytical expression for the polarization of the $c$-$\bar{c}$ pair based on the Fokker-Planck equation under the consideration of spin-vorticity coupling. These polarized $c$-$\bar{c}$ pairs lead to the formation of the $\mathrm{J}/\psi$ at the hadronization surface via the coalescence and fragmentation mechanisms. Correspondingly, we obtain the $m=0$ diagonal element $\rho_{00}$ of the $\mathrm{J}/\psi$ spin-density matrix, which quantifies its spin alignment along the chosen quantization axis. Our study provides a microscopic description of heavy-quark spin transport and suggests rotational diffusion as a possible mechanism underlying the observed $\mathrm{J}/\psi$ spin alignment.
\end{abstract}
\date{\today}
\maketitle

\section{Introduction} 

The heavy quarks (charm and bottom) produced in ultra-relativistic heavy-ion collisions, due to their unique intrinsic characteristics, serve as one of the cleaner probes for the possible existence of deconfined QCD matter. In such collisions, heavy quarks are predominantly produced during the initial hard partonic scatterings and, as they traverse the medium, provide a glimpse of the medium properties. Heavy quarks, produced predominantly in the initial hard scatterings, are sensitive to electromagnetic fields generated in the early stages as well as to the vortical motion of the evolving QCD medium. The dynamics of heavy quarks in the deconfined QCD matter, popularly known as quark-gluon plasma (QGP), have been studied through various observables~\cite{ALICE:2020iev, ALICE:2020pvw, ALICE:2023jad, ALICE:2025cdf, STAR:2026puo, Akamatsu:2008ge, Banerjee:2011ra, vanHees:2007me, Dong:2019unq, ALICE:2023gco, Sahoo:2025kur, Sahoo:2025bkx, Sahoo:2023oid}. Among these, a key aspect of such studies is investigating transport properties, especially estimating the drag and diffusion coefficients that characterize how heavy quarks interact with the surrounding medium~\cite{Akamatsu:2008ge, Banerjee:2011ra, vanHees:2007me, Dong:2019unq}. These studies primarily focus on the translational Brownian motion of heavy quarks, in which medium-induced drag and diffusion drive the evolution of their linear momentum and spatial transport. However, the recent observation of hadron spin polarization at relativistic heavy ion collisions~\cite{ALICE:2019aid, ALICE:2020iev, ALICE:2023jad, ALICE:2025cdf, STAR:2026puo, STAR:2022fan} motivates the investigation of the rotational Brownian motion and spin dynamics of heavy quarks as they propagate through the QGP medium~\cite{Dey:2025ail, Jaiswal:2026ixt, Jaiswal:2026juz, Dey:2026epy}. \\

Rotational Brownian motion describes the stochastic evolution of the orientation (equivalently, the angular momentum) of a particle subjected to a fluctuating torque exerted by a thermal environment, in direct analogy with the fluctuating force responsible for translational Brownian motion. For a rigid body of moment-of-inertia tensor $\mathbf{I}$, the angular velocity $\boldsymbol{\omega}(\tau)$ obeys the Langevin equation
\begin{equation}
\mathbf{I}\,\frac{d\boldsymbol{\omega}(\tau)}{d\tau} = -\boldsymbol{\zeta}_{\rm r}\,\boldsymbol{\omega}(\tau) + \boldsymbol{\xi}(\tau),
\label{eq:rot_langevin}
\end{equation}
where $\boldsymbol{\zeta}_{\rm r}$ is the rotational friction tensor and $\boldsymbol{\xi}(\tau)$ is a stochastic torque, taken to be a Gaussian white noise with zero mean, $\langle \xi_i(\tau)\rangle = 0$, and second moment fixed by the fluctuation--dissipation theorem.\\

In the overdamped (high-friction) regime, inertial effects relax on a timescale $I/\zeta_{\rm r}$ negligible compared with the timescale of orientational decorrelation, and the orientation vector $\hat{\mathbf{u}}(\tau)$ executes diffusion, governed by the rotational diffusion equation;
\begin{equation}
\frac{\partial P(\hat{\mathbf{u}}, \tau)}{\partial \tau} = D_{\rm r}\,\nabla^2_{\hat{\mathbf{u}}}\,P(\hat{\mathbf{u}}, \tau),
\label{eq:rot_diffusion_eq}
\end{equation}
where $\nabla^2_{\hat{\mathbf{u}}}$ is the angular part of the Laplacian and $D_{\rm r} = k_{\rm B}T/\zeta_{\rm r}$ is the rotational diffusion coefficient, related to $\zeta_{\rm r}$ through the rotational Einstein--Stokes--Debye relation. The solution of Eq.~\eqref{eq:rot_diffusion_eq} yields the well-known exponential decay of the orientational autocorrelation function,
\begin{equation}
\langle \hat{\mathbf{u}}(\tau)\cdot\hat{\mathbf{u}}(0)\rangle = e^{-2D_{\rm r}\tau},
\label{eq:orient_corr}
\end{equation}
with the associated orientational relaxation time $\tau_{\rm r} = (2D_{\rm r})^{-1}$. This formalism, originally developed by Debye~\cite{Debye} and later placed on a rigorous statistical-mechanical footing by Chandrasekhar~\cite{Chandrasekhar:1943ws} and Risken~\cite{Risken, Risken:1996}, has since found broad application across statistical physics~\cite{Risken, Risken:1996, ValaKrishnan, Chandrasekhar:1943ws, Matevosyan, Livi, Garcia}, soft matter physics~\cite{Kalmykov, Marbach}, biophysics~\cite{Tirado, GarciadeLaTorre2003}, molecular biology~\cite{Weber, Ermak1978, Woessner1962}, chemical science~\cite{Kojro}, condensed matter systems~\cite{Debye, Brown, Perrin, García-Palacios, Taniguchi, Valiev, Schubert, Nishino}, nanoscience~\cite{Rings}, astrophysics~\cite{Chandrasekhar:1943ws, Merritt:2001dc, Atmaja:2012jg}, and plasma physics~\cite{Waldron}.\\

The translational Brownian motion of a heavy quark propagating through the quark--gluon plasma (QGP) is a well-established paradigm because the heavy-quark mass greatly exceeds the thermal scale of the medium ($M_{\rm Q} \gg T$), momentum transfer per collision with the thermal medium constituents is small, $\Delta p \ll p$. The multiple soft scatterings with thermal partons drive the heavy-quark momentum distribution toward a Fokker--Planck description characterized by a drag coefficient and a momentum-space diffusion coefficient, which are computable from the underlying in-medium interactions. Motivated by the separation of heavy-quark spin-relaxation scales from microscopic interaction scales, we model the heavy-quark spin as a slow degree of freedom undergoing stochastic rotational dynamics induced by its interactions with the QCD medium. \\

Successive stochastic kicks exerted by the medium on the heavy-quark spin vector are modeled by a Langevin equation of the form of Eq.~\eqref{eq:rot_langevin}, with $\boldsymbol{\omega}\to\mathbf{s}/\hbar$. We introduce an effective spin-relaxation coefficient $\zeta_{\rm s}$ and rotational spin-diffusion coefficient $D_{\rm s}$, and adopt the phenomenological Einstein-like relation $D_{\rm s}=k_{\rm B}T/\zeta_{\rm s}$. The consequence is that an initially polarized heavy quark undergoes depolarization as per Eq.~\eqref{eq:orient_corr}, with $D_{\rm r} \to D_{\rm s}$, over a characteristic spin-relaxation time $\tau_{\rm s} \propto (D_{\rm s})^{-1}$. The spin-diffusion coefficient is expected to be related to the same underlying in-medium color-field correlations that govern heavy-quark momentum transport, although their quantitative relation is not established here. The spin and momentum diffusion coefficients of a heavy quark are expected to be correlated through the underlying in-medium interactions, although their precise relation is not established here. Such a connection could provide a useful link between heavy-quark energy-loss and momentum-diffusion phenomenology and spin-sensitive observables such as quarkonium spin alignment and hyperon polarization. This connection between rotational Brownian motion and heavy-quark spin dynamics in the QGP has recently begun to be explored~\cite{Dey:2025ail, Li:2025ugb}.\\

Recently, rotational Brownian motion has been applied to heavy-quark spin dynamics in external magnetic~\cite{Dey:2025ail} and vortical~\cite{Li:2025ugb} fields, with the resulting heavy-quark polarization used to study $D^{*+}$ meson spin alignment~\cite{Dey:2025ail}. However, in deconfined QCD matter, the dynamics of open-charm hadrons and hidden-charm hadrons, such as $J/\psi$, are entirely different~\cite{crs1, crs2, crs3}. In open-charm hadrons, the heavy-quark polarization provides an important contribution to the hadron spin alignment, whereas for hidden charm both the charm and anticharm polarizations enter the $J/\psi$ spin density matrix. In view of these, the present work investigates the rotational Brownian motion of heavy quarks and antiquarks in the QCD medium by employing a semiclassical stochastic Landau–Lifshitz (LL)-like equations, which provide an effective description of the spin-polarization dynamics of fermions propagating through a rotating magnetized medium. Solving the corresponding Fokker-Planck equation perturbatively, we derive an analytical expression for the time evolution of the heavy-quark spin polarization. \\

For the purpose of estimating the maximal surviving polarization, we take the initially produced heavy quarks to be fully polarized along the magnetic-field direction, which is also aligned with the vorticity field generated in non-central relativistic heavy-ion collisions. The residual polarization retained by the heavy quark after undergoing multiple stochastic interactions with the surrounding medium is then estimated within this framework. The spin alignment of the $\mathrm{J}/\psi$ is subsequently obtained by combining the polarizations of the constituent heavy quark and antiquark at the hadronization surface via both coalescence and fragmentation mechanisms. We find that the resulting spin-density matrix element $\rho_{00}$ depends sensitively on the hadronization mechanism, yielding $\rho_{00} > 1/3$ for coalescence and $\rho_{00} < 1/3$ for fragmentation. This study provides an effective framework connecting heavy-quark spin dynamics to the rotational and electromagnetic properties of the QGP medium, offering new insights into spin transport in hot QCD matter. \\

The remainder of this paper is organized as follows. Section~\ref{form} presents the formalism underlying our calculation of heavy-quark rotational Brownian motion and spin polarization. Section~\ref{res} presents and discusses the numerical results, including the residual heavy-quark polarization and the resulting $\mathrm{J}/\psi$ spin alignment. Finally, Section~\ref{sum} summarizes our main findings and their implications.


\section{Formulation}
\label{form}

This section outlines the theoretical framework used in this study. It starts by extending the standard Landau-Lifshitz-Gilbert (LLG) equation--which describes magnetization and spin polarization of particles subjected to a magnetic field--to the spin dynamics of heavy quarks influenced by a vorticity field $\boldsymbol{\omega}$. The generalized LLG equation can be recast as a multivariate Langevin equation, which in turn yields the corresponding Fokker–Planck equation. The solution to the Fokker–Planck equation in the presence of a vorticity field is obtained via a perturbative series expansion. This approach allows determination of the average angular distribution of heavy quarks relative to the vorticity axis, serving as a proxy for heavy-quark or antiquark polarization. 

\subsection{Rotational Brownian motion of heavy quarks}
The stochastic generalized LLG equation, which governs the spin dynamics of a heavy quark in its rest frame under the combined influence of rotation $\boldsymbol{\omega}$ and magnetic field $\boldsymbol{B}$, expressed in terms of $\bf{\Theta}$, is given as follows;
\begin{equation}\label{SpinLangevin}
\frac{d\boldsymbol{s}}{d\tau} = \boldsymbol{s} \times \left[\boldsymbol{\Theta}  + \boldsymbol{\xi}(\tau) \right] - \lambda\, \boldsymbol{s} \times \left( \boldsymbol{s} \times \boldsymbol{\Theta}   \right).
\end{equation}
Here, $\boldsymbol{s}$ represents the classical spin vector, and $\tau$ is the proper time measured in the heavy quark's rest frame. The field associated with the time-dependent spin Hamiltonian $\mathcal{H}(\boldsymbol{s})$ is the combined effect due to the vorticity field and magnetic field, given as $\bf{\Theta} = \boldsymbol{\omega} + \boldsymbol{\tilde{B}} = -\frac{\partial \mathcal{H}}{\partial \boldsymbol{s}}$. Where, $\boldsymbol{\tilde{B}} \equiv \mathcal\gamma \boldsymbol{B}$ and $\gamma$ is the gyromagnetic ratio which connect the magnetic moment $\boldsymbol{\mu}$ with the spin vector $\boldsymbol{s}$ with the relation $\boldsymbol{\mu} = \gamma \boldsymbol{s}$. The noise factor $\boldsymbol{\xi}(\tau)$ is the random torque experienced by the heavy quark due to its interaction with the medium. The term $\boldsymbol{s} \times \boldsymbol{\Theta}$ measures the precession dynamics of the system. The term  $\boldsymbol{s} \times \left(\boldsymbol{s} \times \boldsymbol{\Theta} \right)$ describes the damping mechanism of the system. The factor $\lambda$ is called the damping coefficient, which determines the relative dominance between relaxation and precession dynamics. \\

Equation~\eqref{SpinLangevin} is a special case of the generalized multivariate Langevin equation~\cite{Risken:1996, ValaKrishnan},
\begin{equation}\label{mult_langevin}
\frac{d y_i}{d\tau} = A_i(y, \tau) + C_{ik}(y, \tau) \, \xi_k(\tau),
\end{equation}
where $\xi_k(\tau)$ denotes the noise term.  The stochastic properties of the fluctuating field $\xi(\tau)$ follow the standard correlation rules~\cite{Nishino, Taniguchi};
\begin{equation}\label{Correlation}
\langle \xi_k(\tau) \rangle = 0, \quad
\langle \xi_k(\tau_1)\, \xi_{l}(\tau_2) \rangle = 2\, D_s\, \delta_{kl} \,\delta(\tau_1 - \tau_2),
\end{equation}
which implies that the fluctuations are statistically independent. Here, $D_s$ is the spin diffusion coefficient. Using the Kramers–Moyal expansion for Eqs.~\eqref{mult_langevin} and \eqref{Correlation}, one can obtain the Fokker–Planck equation~\cite{Risken:1996, Livi};
\begin{align}
\frac{\partial \mathcal{P}}{\partial \tau} = &\,
- \frac{\partial}{\partial y_i} 
\left[ A_i(y, \tau) + D_{s} \, C_{jk}(y, \tau) 
\frac{\partial C_{ik}(y, \tau)}{\partial y_j} \right] \mathcal{P} \nonumber\\
&\,+ D_{s} \, \frac{\partial^2}{\partial y_i \partial y_j} 
\left[ C_{ik}(y, \tau) C_{jk}(y, \tau) \mathcal{P} \right],
\label{eq:Genfokker_planck}
\end{align}
where, $\mathcal{P} \equiv\mathcal{P}(y,\tau|y_0,\tau_0)$ is the transition probability from the state $(y_0,\tau_0)$ to $(y,\tau)$. It is worth mentioning that Eq.~\eqref{eq:Genfokker_planck} is determined by the coefficients of the Langevin equation.\\

Next, Eq.~\eqref{SpinLangevin} can be expressed in terms of a general multivariate Langevin equation (Eq.~\eqref{mult_langevin}) by identifying $s_i=y_i$ and 
\begin{align}
A_i &=  \epsilon_{ijk} s_j \Theta_k + \lambda\, (s^2 \delta_{ik} - s_i s_k) \Theta_k, \label{eq:A_coefficient} \\
C_{ik} &= \epsilon_{ijk} s_j. \label{eq:B_coefficient}
\end{align}
Now, for obtaining the aforementioned relations, we have used the following identities:
\begin{align}
\frac{\partial C_{ik}}{\partial s_j} =&\, \epsilon_{ijk}, \quad  C_{jk} \, \frac{\partial C_{ik}}{\partial s_j} = -2\,s_i,  
\label{eq:diffusion_derivative}\\
C_{ik} C_{jk} =&\, s^2 \delta_{ij} - s_i s_j. \label{CikCjk}
\end{align}
Finally, we get the Fokker–Planck equation corresponding to the stochastic Landau–Lifshitz-Gilbert formulation, which can be written as~\cite{Nishino, García-Palacios, Taniguchi};
\begin{align}
\frac{\partial \mathcal{P}}{\partial \tau} = &\,
-\! \frac{\partial}{\partial s_i} 
\left[
\epsilon_{ijk}\, s_j \Theta_k + \lambda
(s^2 \delta_{ik} - s_i\, s_k) \Theta_k \!
- 2D_{s} s_i \right] \mathcal{P} \nonumber\\
&\, + D_{s} \, \frac{\partial^2}{\partial s_i\, \partial s_j} 
\left[ s^2 \delta_{ij} - s_i s_j \right] \mathcal{{P}}, \label{delPdelt}
\end{align}
where $\mathcal{P}\equiv \mathcal{P}(\boldsymbol{s},\tau)$ is the probability distribution for spin orientation along $\textbf{s}$ at time $\tau$.\\

In this study, we have considered the vorticity field $\boldsymbol{\omega}$ to be independent of particle spin $\boldsymbol{s}$. Thus, the Eq.~\eqref{delPdelt} reduce to a Smoluchowski-like equation~\cite{Garcia},
\begin{equation}
\frac{\partial \mathcal{P}}{\partial \tau} =
D_{s}\, \frac{\partial}{\partial \boldsymbol{s}} \cdot 
\left[
\boldsymbol{s} \times 
\left(
\boldsymbol{s} \times 
\left( \frac{\lambda}{D_{s}} \boldsymbol{\Theta} - \frac{\partial}{\partial \boldsymbol{s}} \right)
\right)
\right] \mathcal{P},
\label{eq:smoluchowski}
\end{equation}
which is the rotating counterpart to the Klein-Kramers equation~\cite{Chandrasekhar:1943ws}. The instantaneous spin orientation of a spin-polarized particle along the direction $(\theta,\phi)$ is described in terms of a sphere of fixed radius $s$ in spin space, with $\boldsymbol{s} = (s,\theta,\phi)$ in spherical coordinates. It is defined in such a way that each point on the sphere corresponds to a distinct spin orientation of the particle~\cite{Kalmykov, Narducci}. Here, we consider the direction of the vorticity field $\boldsymbol{\omega}$ to be along the $z$-direction. \\

Assuming an axially symmetric Hamiltonian, Eq.~\eqref{eq:smoluchowski} simplifies as~\cite{Kalmykov, Debye, Brown};
\begin{equation} \label{eq:spherical_fokker_planck}
\tau_s \frac{\partial \mathcal{P}}{\partial \tau} =
\frac{1}{\sin \theta} \frac{\partial}{\partial \theta} 
\left[
\sin \theta 
\left(\frac{\lambda}{D_{s}}
\frac{\partial \mathcal{H}}{\partial \theta} \mathcal{P} +  \frac{\partial \mathcal{P}}{\partial \theta}
\right)
\right]
\end{equation}
where, $\tau_s\equiv 1/D_s$ is the spin relaxation time
and $\mathcal{H} = - \boldsymbol{\omega}\cdot\boldsymbol{s} - \boldsymbol{\mu} \cdot \boldsymbol{B} = - (\boldsymbol{\omega} + \boldsymbol{\tilde{B}}) \cdot \boldsymbol{s}$. In the Hamiltonian, the first term $\boldsymbol{\omega}\cdot\boldsymbol{s}$ represents the spin-vorticity coupling, and the second term $\boldsymbol{\mu} \cdot \boldsymbol{B}$ denotes the spin-magnetic coupling. Here, it is important to mention that spin polarization is obtained in the particle's rest frame. Thus, the vorticity and magnetic field are calculated in the rest frame of the particle using the Lorentz transformation;
\begin{align}
\boldsymbol{\omega} &= \gamma_{v}\, \left( \boldsymbol{\omega_{\rm Lab}} - \boldsymbol{k_{\rm Lab}}\times\boldsymbol{v} \right) + \left( 1 - \gamma_{v} \right) \left( \frac{\boldsymbol{\omega_{\rm Lab}} \cdot \boldsymbol{v}}{|\boldsymbol{v}|^2} \right) \boldsymbol{v} ,\label{Lorentz_trans_om}\\
\boldsymbol{B} &= \gamma_{v}\, (\boldsymbol{B_{\rm Lab}} - \boldsymbol{E_{\rm Lab}} \times \boldsymbol{v} ) + \left( 1 - \gamma_{v} \right) \left( \frac{\boldsymbol{B_{\rm Lab}} \cdot \boldsymbol{v}}{|\boldsymbol{v}|^2} \right) \boldsymbol{v}, \label{Lorentz_trans_B}
\end{align}
where $\gamma_{v} = 1/\sqrt{1-\boldsymbol{v^{2}}} \equiv E/m_Q$ is the Lorentz gamma factor. Here, $\boldsymbol{v}$ is the velocity of the heavy quark, and $\boldsymbol{E_{\rm Lab}}$, and $\boldsymbol{B_{\rm Lab}}$ are the laboratory frame electric and magnetic fields, respectively. Similarly, in Eq.~(\ref{Lorentz_trans_om}) $\boldsymbol{\omega_{\rm Lab}}$ is the vorticity in the laboratory frame and $\boldsymbol{k_{\rm Lab}}$ is a vector obtained by projecting the spin-polarization tensor along the fluid velocity in the laboratory frame. \\

A general definition of the four-vector $k_{\mu}$ associated with the spin polarization tensor is given in Eq.~(\ref{adeq2}). In our calculation, we have used the ensemble-averaged values of the vorticity, Eq.~(\ref{Lorentz_trans_om}), and magnetic field, Eq.~(\ref{Lorentz_trans_B}). A brief discussion of these variables is provided in Appendix~\ref{appexA}.
From Eq.~(\ref{Lorentz_trans_om}) and (\ref{Lorentz_trans_B}), one may extract that for a given $p_{\rm T}$ value, the vorticity field and magnetic field in the heavy quark's rest frame increase with rapidity. Therefore, the signal of spin polarization induced by the initial magnetic field is expected to be more prominent for heavy quarks with finite rapidity due to the enhancement of the magnetic field in the particle's rest frame.\\

The antisymmetric spin polarization tensor $\omega_{\mu\nu}$ is decomposed as~\cite{Florkowski:2017ruc}
\begin{equation}
\omega_{\mu\nu} = k_{\mu}u_{\nu} - k_{\nu}u_{\mu} + \epsilon_{\mu\nu\alpha\beta}u^{\alpha}\omega^{\beta}
\label{adeq1}
\end{equation}
where, $k_{\mu}$ and $\omega_{\mu}$ are defined in terms of spin polarization tensor
\begin{equation}
k_{\mu} = \omega_{\mu\nu}u^{\nu}, \qquad \omega_{\mu} = \frac{1}{2} \epsilon_{\mu\nu\alpha\beta}\omega^{\nu\alpha}u^{\beta}
\label{adeq2}
\end{equation}
To hold the relation $k^{\mu}u_{\mu} = \omega^{\mu}u_{\mu} = 0$, the $\omega_{\mu}$ and  $k_{\mu}$ are set to orthogonal to the fluid velocity $u_{\mu}$. Here, $\epsilon_{\mu\nu\alpha\beta} $ is the Levi Civita antisymmetric four tensor, $\epsilon^{0123}= - \epsilon_{0123} = 1$. In this work, we consider a rigid rotation of the fluid around the $y$ axis, i.e., $\hat{\omega}_{\mu} = (0,0,1, 0)$. We have considered the velocity profile  $ u^{\mu} =\gamma_{\omega}(1, \Omega z, 0, -\Omega x)$, with $\gamma_{\omega} = 1/\sqrt{1-\Omega^{2}r^2}$, here $r$ is the distance from the center of the vortex in the transverse plane, $r^2 = x^2 + z^2$. In this framework, $\boldsymbol{\omega}$ is defined via the relation $\omega^\mu=(0,\,0,\,\boldsymbol{\omega}/T,\,0)$.\\

One needs to solve Eq.~(\ref{adeq1}) and ~(\ref{adeq2}) self-consistently to obtain the spin polarization tensor, $\omega_{\mu\nu}$; 
\begin{equation}
\omega_{\mu\nu} =
\left[ {\begin{array}{cccc}
0 & 0 & 0 & 0 \\
0 & 0 & 0 & \frac{\Omega}{T_0} \\
0 & 0  & 0 & 0 \\
0 & -\frac{\Omega}{T_0} & 0 & 0 \\
\end{array} } \right], \label{adeq3}
\end{equation}
where the parameter $T_0$ is introduced to keep $\omega_{\mu\nu}$ dimensionless. Now, substituting Eq.~\eqref{adeq3} in Eq.~\eqref{adeq2}, we obtain, $k_{\mu} = (\Omega^2{\gamma_{\omega}/T_{0}})(0,z,0,x)$,\; and $\omega_{\mu} = (\Omega{\gamma_{\omega}/T_{0}})(0,0,1,0)$.\\

Further, substituting these values in Eq.~\eqref{eq:spherical_fokker_planck}, we get
\begin{align}\label{FokPM}
&\tau_s \frac{\partial \mathcal{P}}{\partial \tau} = \nonumber \\ &
\frac{1}{\sin \theta} \frac{\partial}{\partial \theta} \Bigg[ \sin \theta \Bigg(\frac{\partial}{\partial \theta} + \frac{\lambda}{D_s} \left(\frac{\gamma_{\omega} T  s \Omega}{T_{0}} + \mu B(\tau) \right) \sin \theta\, \Bigg) \Bigg]\mathcal{P}. 
\end{align}
The following section presents the solution of Eq.~\eqref{FokPM}, which describes the diffusion rate of heavy quarks in the medium and its impact on spin polarization.

\subsection{Heavy quark polarization} 

We now solve the Fokker--Planck equation governing the spin-polarization distribution $\mathcal{P}(\boldsymbol{s},\tau)$ of heavy quarks, via a perturbative series expansion, from which the mean spin polarization along the vorticity axis is obtained. The Fokker-Planck equation given in Eq.~\eqref{FokPM} can be written in the reduced form as,
\begin{align}\label{fokker_planck}
\tau_s\, \partial_{\tau}\mathcal{P}(\theta,\tau) = \mathcal{L}_{\theta}(\tau) \, \mathcal{P}(\theta,\tau).
\end{align}
Here, $\partial_\tau\equiv\frac{\partial}{\partial \tau}$ and $\mathcal{L}_\theta (\tau)$ is a time-dependent differential operator with respect to $\theta$. Considering the time evolution in operator form,
\begin{equation}
\label{fokpl_ope}
\tau_s\,\partial_{\tau} \ket{\mathcal{P},\tau} = \hat{\mathcal{L}}(\tau) \ket{\mathcal{P},\tau}.
\end{equation}
It is easy to see that the generic solution has the following structure.
\begin{equation}
\ket{\mathcal{P},\tau}=\text{exp}\left[ \frac{1}{\tau_s}\int_{0}^{\tau}d\tau^\prime\, \hat{\mathcal{L}}(\tau^\prime)\right] \ket{\mathcal{P},0}
\end{equation}
We note that the differential operator $\hat{\mathcal{L}}$ can be separated into a time-independent as well as a time-dependent part, i.e., $\hat{\mathcal{L}}^{0}$ and $\hat{\mathcal{L}}^\prime(\tau)$ respectively. \\

For the temporal evolution of the magnetic field, we assumed an exponentially decaying profile, $B(\tau) = B_0 \phi(\tau)$ with $\phi(\tau)=e^{-\tau/\tau_B}$, where $\tau_B$ denotes the characteristic timescale at which the magnetic field approaches zero. Further considering, a constant vorticity profile $\Omega = \Omega_0 $, one can write,
\begin{equation}
\hat{\mathcal{L}}(\tau)=\hat{\mathcal{L}}^{0} + \alpha \hat{\mathcal{L}}^{\prime} +  \beta\hat{\mathcal{L}}^{\prime \prime}(\tau),
\end{equation}
where $\alpha\equiv \frac{\lambda}{D_s} \frac{\gamma_{\omega} T  s \Omega_0}{T_{0}}$, and $\beta \equiv \frac{\lambda \;\mu B_0}{D_s} $. Further, by treating $\alpha$ and $\beta$ as small perturbative parameters, the solution corresponding to the probability distribution for spin orientation is obtained using the Dyson series expansion in powers of $\alpha$ and $\beta$ as follows~\cite{Risken:1996, Chun};
\begin{align}
&\mathcal{P}\,(\theta,\tau;\theta_{0},0)\equiv \bra{\theta}\ket{\mathcal{P},\tau} =\mathcal{P}^{0}(\theta,\tau;\theta_{0},0) \nonumber   \\
& + \alpha \!\!\int_\Omega\!\int_{0}^{t}\!\!\frac{d\tau^\prime}{\tau_s}d\Omega^\prime\mathcal{G}^{0}_{\tau-\tau^{\prime}}(\theta,\theta^\prime)\mathcal{L}^\prime_{\theta^\prime}\mathcal{P}^{0}(\theta^\prime\!,\tau^\prime;\theta_{0},0) \!+\! \mathcal{O}(\alpha^2) \nonumber   \\
& +\beta \!\!\int_\Omega\!\int_{0}^{t}\!\!\frac{d\tau^\prime}{\tau_s}d\Omega^\prime\mathcal{G}^{0}_{\tau-\tau^{\prime}}(\theta,\theta^\prime)\mathcal{L}^{\prime\prime}_{\theta^\prime}(\tau^\prime)\mathcal{P}^{0}(\theta^\prime\!,\tau^\prime;\theta_{0},0) \!+\! \mathcal{O}(\beta^2) \, ,
\label{SolDyson}
\end{align}
where, $d\Omega^\prime \equiv 2\pi \sin\theta^\prime\,d\theta^\prime$.\\ 

In Eq.~\eqref{SolDyson}, $\mathcal{P}^{0}(\theta,\tau;\theta^\prime,0)$ and $\mathcal{G}_{\tau}^{0}(\theta,\theta^\prime)$ are the solution and the corresponding Green function of Eq.~\eqref{fokker_planck}, respectively, at $\alpha=0$, and $\beta$ = 0. In the absence of vorticity and magnetic field, the operator $\hat{\mathcal{L}}$ in Eq.~\eqref{fokpl_ope} consists only of the time-independent part $\hat{\mathcal{L}}^{0}$, which is given by
\begin{equation}
\hat{\mathcal{L}}^{0} = \frac{1}{\sin \theta} \frac{\partial}{\partial \theta} \bigg( \sin \theta \frac{\partial}{\partial \theta} \bigg).
\end{equation}
In this case, solution of Eq.\,\eqref{fokker_planck} can be expressed as 
\begin{align}
\mathcal{P}(\theta,\tau ;\theta^\prime,0)=\int d\Omega^\prime \, \mathcal{G}_{\tau}(\theta,\theta^\prime) \, \mathcal{P}(\theta^\prime,0)\, ,
\label{Ptheta}
\end{align}
where,
\begin{align}
\mathcal{G}_{\tau}(\theta,\theta^\prime)=\sum_{n=0}^{\infty}\frac{(2n+1)}{4\pi}\exp\left[-n(n+1)\frac{\tau}{\tau_s}\right]\nonumber\\
\times P_{n}(\cos\theta) \, P_{n}(\cos\theta^\prime)\, .
\end{align}
Here, $P_{n}(x)$ are Legendre polynomials in $x$ of degree $n$. \\

For this study, we assume that all heavy quarks are initially spin-polarized along the direction of the magnetic field, i.e., $\theta=\theta_{0}$; the respective probability distribution can be written as
\begin{equation}
\mathcal{P}(\theta,0) = \frac{1}{2\pi}\delta(\cos\theta-\cos\theta_{0}) \, .
\end{equation}
Using this condition in Eq.~\eqref{Ptheta}, the solution of Eq.~\eqref{fokker_planck} reduces to 
\begin{align}
\mathcal{P}(\theta,\tau ;\theta_{0},0) = \sum_{n=0}^{\infty}\frac{(2n+1)}{4\pi}\exp\left[-n(n+1)\frac{\tau}{\tau_s}\right]\nonumber\\
\times P_{n}(\cos\theta)P_{n}(\cos\theta_{0})\, .
\end{align}
The solution obtained in the absence of the vorticity (or magnetic) field serves as the leading-order solution for the perturbative treatment.\\ 

The differential operators $\mathcal{L}^\prime$ is given by
\begin{align}
\mathcal{L}^\prime=\frac{1}{\sin\theta}\frac{\partial}{\partial\theta}\,\sin^{2}\theta\,,
\end{align}
and the time-dependent differential operators $\mathcal{L}^\prime(\tau)$ is given by
\begin{align}
\mathcal{L}^{\prime\prime}(\tau)=\frac{\phi(\tau)}{\sin\theta}\frac{\partial}{\partial\theta}\,\sin^{2}\theta\, .
\end{align}
Employing the expression given in Eq.~\eqref{SolDyson}, one can directly calculate the expectation value of $\cos\theta$ as 
\begin{align}
&\langle \cos\theta \rangle  = \cos\theta_{0} \, e^{- 2\tau/\tau_s} \nonumber \\ & + \frac{2\alpha}{3\tau_s }\, e^{-2\tau/\tau_s} \int_{0}^{\tau} \! d\tau^\prime \, 
\Big[e^{2\tau^\prime/\tau_s} - P_{2}(\cos\theta_{0}) e^{-4\tau^\prime/\tau_s}\Big]\, \nonumber \\ + & \frac{2\beta}{3\tau_s}\, e^{-2\tau/\tau_s}  \int_{0}^{\tau} \! \phi(\tau^{\prime}) d\tau^\prime \, \Big[e^{2\tau^\prime/\tau_s} - P_{2}(\cos\theta_{0}) e^{-4\tau^\prime/\tau_s}\Big]\, .
\label{timPol}
\end{align}
It is possible to perform the integrals in the above equation with certain physics-motivated assumptions, as outlined below. \\   

The magnetic field generated in relativistic heavy-ion collisions is expected to be short-lived and to decay rapidly, primarily due to the relativistic motion of the spectator nucleons away from the collision zone. Thus, the magnetic-field strength decreases exponentially with time. Furthermore, although the initial orbital angular momentum generated in a relativistic heavy-ion collision is extremely large, the lifetime of the produced system is very short, typically of the order of a few fm/$c$. Consequently, the total angular displacement of the system during its evolution is expected to be infinitesimal. Therefore, the medium produced in heavy-ion collisions does not undergo macroscopic rotation in the conventional sense. In our treatment, we consequently assume that the vorticity decreases with a constant characteristic decay timescale during the evolution of the system. Thus, by employing the aforementioned magnetic decay profile for a rigid-like rotation of the system, one can perform the integral in Eq.~\eqref{timPol} to obtain $ \langle \cos\theta \rangle$ as;
%
\begin{align}
\langle \cos\theta \rangle &=\cos\theta_{0} \, e^{- 2\tau/\tau_s} \nonumber\\ 
&+ \frac{\alpha}{3} \! \left[(1-e^{-2\tau/\tau_s}) + \frac{1}{2} (e^{-6\tau/\tau_s}-e^{-2\tau/\tau_s}) \right] \nonumber \\ 
&+ \frac{2\beta \tau_B}{3} \; e^{-2\tau/\tau_s} \nonumber \\
&\times\left[\frac{1-e^{-(\tau_s - 2\tau_B)\tau/\tau_s \tau_B}}{\tau_s - 2\tau_B} -\frac{1-e^{-(\tau_s + 4\tau_B)\tau/\tau_s \tau_B}}{\tau_s + 4\tau_B} \right].
\label{eq:costheta}
\end{align}
%
It is worth mentioning that to obtain the solution for polarization, we have used $P_{2n}(\cos\theta_{0})=1$ and $P_{2n+1}(\cos\theta_{0})=\cos\theta_{0}$, with $\theta_{0}=0,\pi$.\\

The quantity $\langle \cos\theta \rangle$ measures the heavy quark polarization induced by the vorticity field. In the above equations, $\tau$ is the duration for which the heavy quark undergoes Brownian motion within the QGP, and it is given by $\tau = \frac{Rm_Q}{|\bf{p}|}$. Here, $R$ is the freeze-out radius of the fireball, $m_Q$ is the heavy quark mass, and $\bf{p}$ = $\sqrt{p^2_{\rm T} \;\rm cosh^2 y + m^2_{\rm Q} \; \rm sinh^2 y}$ is the heavy-quark momentum.\\

To validate the perturbative treatment of the constant vorticity field and time-dependent magnetic field, we estimate the maximum value of $\alpha$ and $\beta$. The perturbation parameters $\alpha$ and $\beta$ can be estimated as\footnote{In the absence of an Einstein-Stokes-like relation between drag ad diffusion coefficients for spin evolution, we assume a similar relation based on dimensional arguments, i.e., $D_{s}=\lambda T$.}, 
\begin{equation}
\alpha \equiv  \frac{s \; \gamma_{\omega} \Omega \lambda T }{D_s T_0} \sim \frac{s\; \gamma_{\omega} \Omega}{T_0} , \quad \;\beta \equiv \frac{\lambda \;\mu B}{D_s} \sim \frac{g\;s\;q \gamma_v B}{2m_Q T} \label{alpha}
\end{equation}
where, $\mu = \frac{gsq}{2m_Q} $ is the magnetic moment. Here, $g$ is the g-factor and is taken to be 2, the spin $s = \hbar/2$, and charge $q = fe$, with $f$ = 2/3 for the charm quark. Equation~\eqref{alpha} has several interesting features. It indicates that the perturbation parameter $\alpha$ depends on $\gamma_{v} \equiv E/m_Q = (m_{\rm T} \cosh y)/m_{Q}$,  which is a function of rapidity. Consequently, the results depend on the choice of the rapidity window. Since $\cosh y$ has a diverging nature at higher $y$, the value of $\alpha$ increases as a function of $p_{\rm T}$ and $y$. Therefore, the present study is restricted to the midrapidity region. At mid rapidity ($|y| <$ 1.0), the maximum value of perturbation parameters $\alpha$ is  $0.072$ and $\beta$ = 0.02 for charm quarks with transverse momentum $20$~GeV/c. To obtain the $\alpha$ and $\beta$ values, we use the average QGP temperature of $300$~MeV. The vorticity field and magnetic field in the lab frame are along the $y$-direction and typically take the values $\Omega_{\rm Lab}\sim 0.02$ fm$^{-1}$ and $B_{\rm Lab}$ = 0.1 $m_{\pi}^{2}$, respectively. These values justify treating $\alpha$ and $\beta$ as perturbation parameters, validating the perturbative approach followed earlier. \\


\subsection{Spin-alignment of quarkonia}
\label{quarkonia}
The spin alignment of the vector mesons is studied via the elements of the 3 $\times$ 3 Hermitian spin density matrix ($\rho_{m,m^{'}}$), where $ m $ and $m^{'}$ label the spin component along the quantization axis. Out of these three diagonal elements, $\rho_{00}$ is independent and can be measured in experiments in two-body decays to pseudoscalar mesons or fermions. The spin density matrix element $\rho_{00}$ quantifies the probability of the vector meson being in a spin-zero projection along the quantization axis.\\ 

The spin alignment of vector mesons depends on the hadronization mechanisms~\cite{Liang:2004xn}. Mainly, there are two dominant mechanisms for hadronizing quark-antiquark pairs in different transverse-momentum regions: coalescence (recombination) and fragmentation. 

\begin{enumerate}
\item Coalescence: In the quark coalescence model, the $\rho_{00}$ element is given by~\cite{Liang:2004xn},

\begin{equation}
\rho_{00}^{(\rm Coal)} = \frac{1-P_q P_{\bar{q}}}{3+P_q P_{\bar{q}}}
\end{equation}
where, $P_q (P_{\bar{q}})$ is the polarization of quark (anti-quark).  The polarization of heavy quarks can be obtained using the relation
\begin{equation}
P_{q} = \langle \rm cos\theta\rangle
\end{equation}
The mathematical expression of $\langle \rm cos\theta\rangle$ is given in Eq.~(\ref{eq:costheta}). 

\item Fragmentation: In the fragmentation process, the $\rho_{00}$ element is given by~\cite{Liang:2004xn},

\begin{equation}
\rho_{00}^{(\rm Frag)} = \frac{1+ \beta_f P_q P_{\bar{q}}}{3- \beta_f P_q P_{\bar{q}}}
\end{equation}
where $\beta_f$ = 0.5.
\end{enumerate}


\section{Results and discussion} 
\label{res}
\begin{figure}[ht!]
\centering
\includegraphics[scale = 0.45]{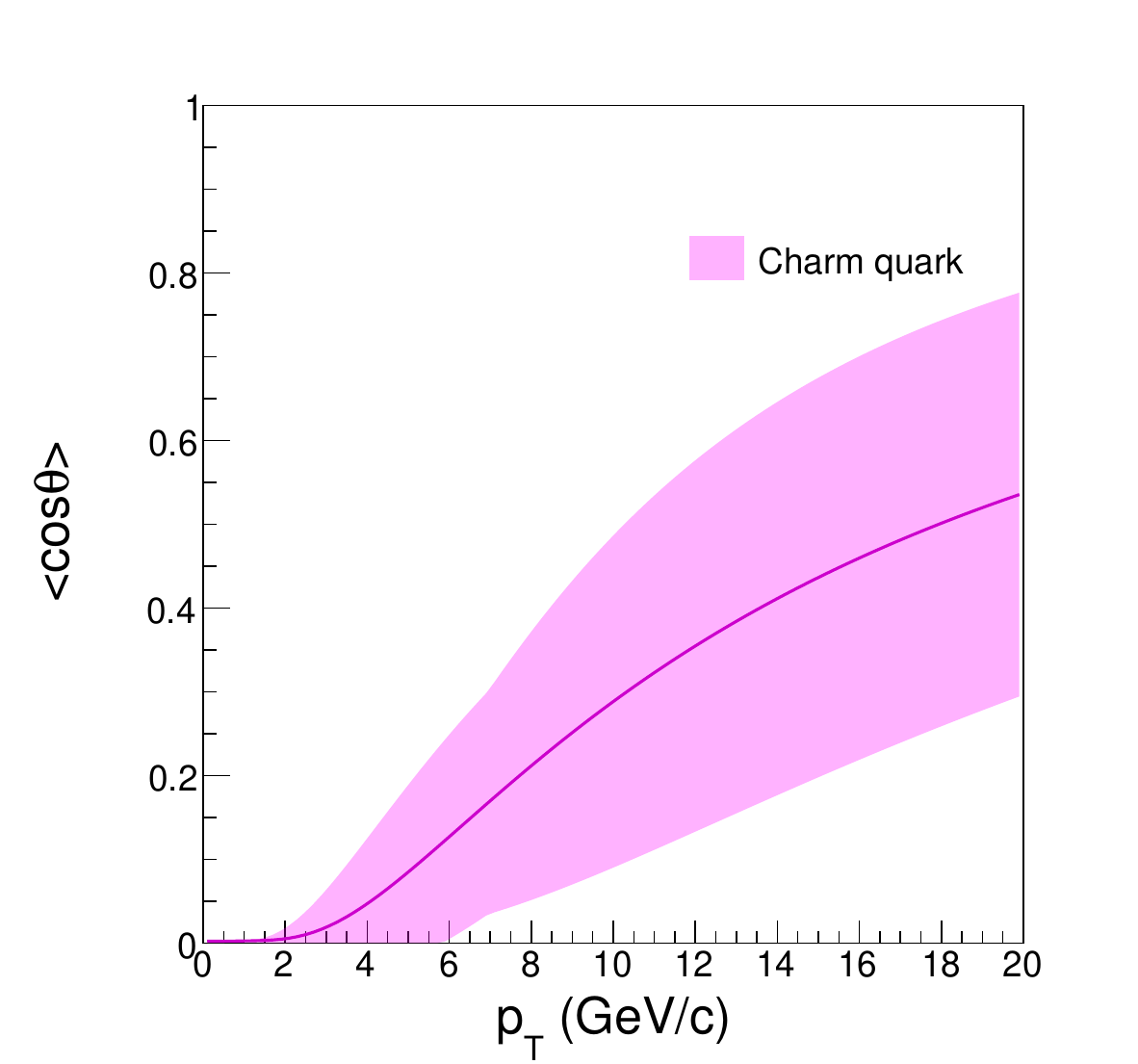}
\caption{The average angular distribution of charm quarks along the direction of the vorticity field as a function of transverse momentum. The systematic band on the theoretical prediction is due to the different values of the spin relaxation time $\tau_s$ used in the model.}
\label{fig:cosTheta}
\end{figure}

We now turn to the spin polarization signal that emerges from vortical charm-quark transport in the QCD medium. Figure~\ref{fig:cosTheta} shows the average angular distribution, $\langle \rm \cos\theta \rangle$, for charm quarks as a function of transverse momentum $p_{\rm T}$, obtained using Eq.~\eqref{timPol}. The quantity $\langle \cos\theta \rangle$ measures the vector polarization of heavy quarks along the local vorticity field, and its deviation from zero quantifies the polarization retained by charm quarks after undergoing interactions with the QCD medium. We find a systematic increase of $\langle \cos\theta \rangle$ with $p_{\rm T}$, indicating that relatively fast-moving charm quarks preserve a larger fraction of the vorticity-induced polarization imprinted during their propagation through the medium. This $p_{\rm T}$-dependence constitutes a distinctive signature of vortical spin transport that can be tested against heavy-flavor hadron polarization measurements. The theoretical uncertainty band associated with this prediction, governed by Eq.~\eqref{eq:costheta}, originates entirely from the spin relaxation time $\tau_{\rm s}$. To estimate this uncertainty, two extreme values, $\tau_s = 1.0$ fm and $\tau_s = 3.0$ fm, are considered. As a first quantitative estimate, we constrain $\tau_{\rm s}$ by fitting the prompt $D^{*+}$ spin-alignment data measured in Pb--Pb collisions at $\sqrt{s_{\rm NN}} = 5.02$~TeV~\cite{ALICE:2025cdf}, which yields $\tau_{\rm s} = 1.31$~fm. However, the preliminary results from the latest Run~3 measurements reveal a pronounced suppression of the high-$p_{\rm T}$ polarization signal relative to Run~2, translating into a substantially different extracted value of $\tau_{\rm s}$. This uncertainty between datasets underscores the need for the broader $\tau_{\rm s}$ range adopted in the present analysis. It highlights spin-alignment observables as a sensitive yet still unsettled probe of charm-quark spin dynamics in the QGP medium.\\

\begin{figure}[ht!]
\centering
\includegraphics[scale = 0.45]{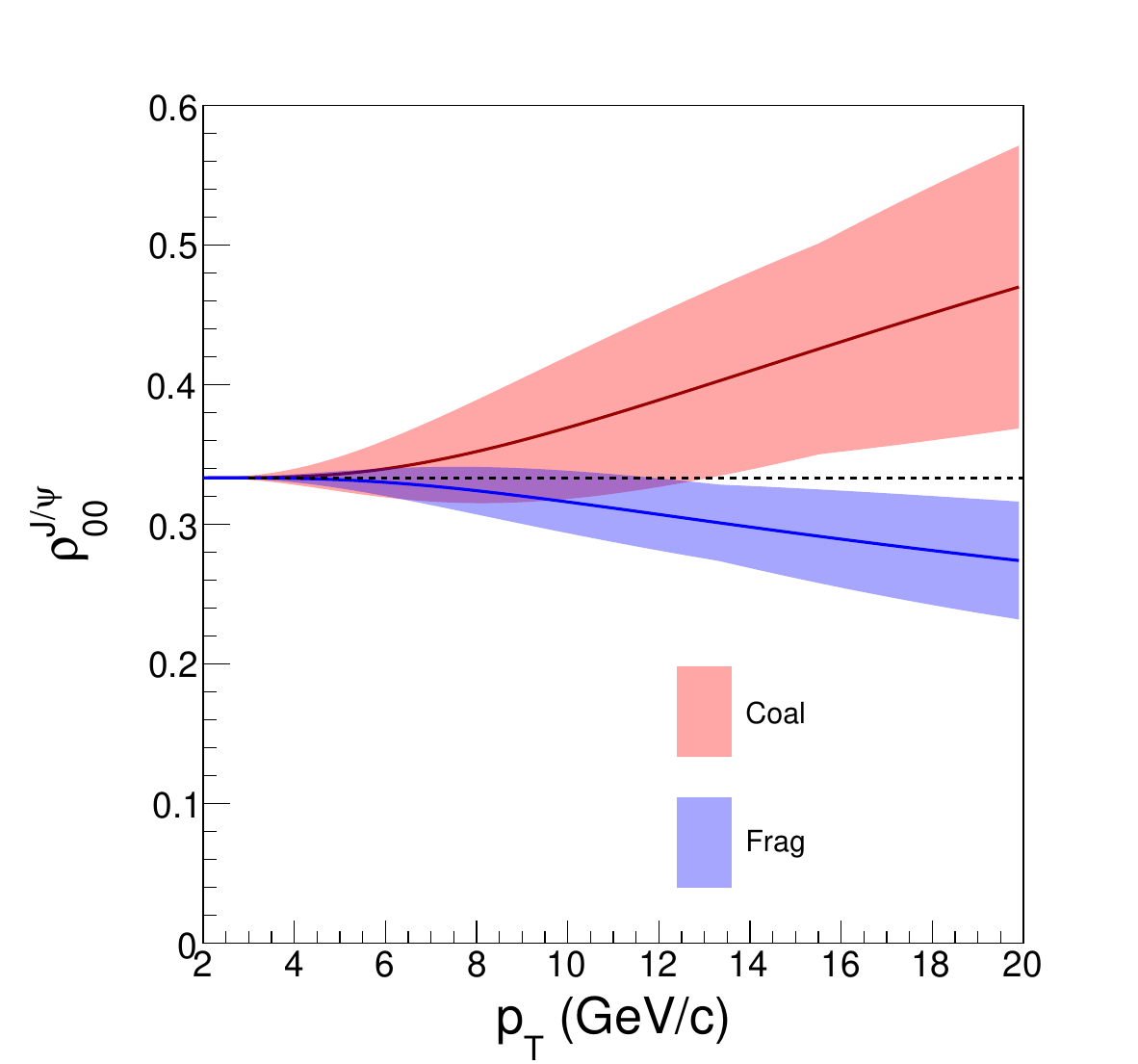}
\caption{The spin alignment is predicted in terms of $00^{\rm th}$ element of the $\mathrm{J}/\psi$ spin-density matrix, $\rho^{\mathrm{J}/\psi}_{00}$, as the function of $p_{\rm T}$ at mid-rapidity for the two hadronization processes: coalescence (Coal) and fragmentation (Frag). The band in the $\rho^{\mathrm{J}/\psi}_{00}$ reflects the sensitivity to the chosen range of spin relaxation time $\tau_s$.} 

\label{fig:Rho00}
\end{figure}

Building on the charm-quark residual polarization, we now predict the spin alignment of $\mathrm{J}/\psi$ at the chemical freeze-out boundary. Figure~\ref{fig:Rho00} shows the $00^{\rm th}$ element of the spin-density matrix, $\rho_{00}$, as a function of transverse momentum $p_{\rm T}$ at mid-rapidity, evaluated for two hadronization mechanisms: coalescence (Coal) and fragmentation (Frag). The result shows that the $\mathrm{J}/\psi$ spin alignment depends strongly on the hadronization mechanisms. For light-flavor vector mesons such as $K^{*0}$, $\phi$, and $\rho$, fragmentation yields $\rho_{00} > 1/3$ while recombination yields $\rho_{00} < 1/3$~\cite{Liang:2004xn}; while in the heavy-flavor sector, we find the opposite hierarchy, with fragmentation giving $\rho_{00} < 1/3$ and recombination giving $\rho_{00} > 1/3$. This reversal originates from the opposite alignment of heavy quarks and antiquarks along the initial magnetic field direction, which we implement via $\cos\theta_0 = +1$ for heavy quarks and $\cos\theta_0 = -1$ for antiquarks. A further prediction of the model is that quarkonium states exhibit stronger spin alignment at high $p_{\rm T}$ because their constituent heavy quarks spend comparatively shorter time interacting with the medium, thereby suppressing spin relaxation.\\

The experimentally measured polarization parameter $\lambda_{\theta}$ is connected to the diagonal element $\rho_{00}$ of the spin-density via the relation;  $\rho_{00} = \frac{1-\lambda_{\theta}}{3+ \lambda_{\theta}}$. If $\lambda_{\theta} <$ 0, then $\rho_{00} > \frac{1}{3}$ and vise-versa. At low-$p_{\rm T}$ ($\leq$ 4 GeV/c), the theoretical model predicts $\rho_{00} = \frac{1}{3}$, indicating no spin alignment. 
This behavior may be attributed to the stronger interaction of low-$p_{\rm T}$ particles with the medium, which leads to a substantial loss of their initial spin polarization through spin relaxation.

\section{Conclusions and outlook} 
\label{sum}
We have explored the rotational Brownian motion of heavy quarks within a QCD medium and demonstrated spin alignment of $\mathrm{J}/\psi$ with the $\rho_{00}$ observable. It is considered that heavy quarks are initially polarized, possibly due to the strong magnetic field aligned with the vorticity field. The loss of initial polarization due to interactions with the QCD medium is obtained using the Fokker–Planck transport equation. Further, the spin alignment of $\mathrm{J}/\psi$ arising from $c-\bar{c}$ is estimated depending on the two formation mechanisms: coalescence and fragmentation. It is observed that the spin alignment of $\mathrm{J}/\psi$ is sensitive to the hadronization mechanisms: $\rho_{00} > \frac{1}{3}$ for coalescence and $\rho_{00} < \frac{1}{3}$ for fragmentation. These findings suggest that $\mathrm{J}/\psi$ exhibits a higher degree of spin alignment at high $p_{\rm T}$, where shorter medium interaction times suppress spin relaxation.
The current study provides insight into heavy-quark spin dynamics in a rotating system, within the Langevin and Fokker-Planck frameworks in angular momentum space.\\

\noindent
Notably, the results presented in this study are based on the model parameters adopted in the analysis. In the absence of first-principles determination of the model parameters at this moment, the results should be regarded as qualitative. A quantitative description of the spin-alignment observable would require a systematic Bayesian inference analysis to constrain the model parameters, such as medium temperature $T$,  freeze-out radius $R$, initial vorticity $\Omega$, spin relaxation time $\tau_s$, magnetic decay parameter $\tau_B$, vorticity decay time $\tau_w$, etc., combined with experimental data. Additionally, the present study focuses on the spin alignment of the only directly produced $\mathrm{J}/\psi$, and it leaves the feed-down contributions from higher-excited states, such as $\psi(2S)$ and $\chi_{c}$, for future study. Moreover, incorporating additional microscopic effects, such as in-medium effects on charmonium dynamics, its regeneration, and possible spin-dependent recombination mechanisms, would enable a more comprehensive but complex description of quarkonium spin alignment throughout the evolution of the medium. Such developments would provide a more realistic connection between the microscopic production and interaction mechanisms of charmonium and the experimentally measured spin observables. Together, these improvements could help establish quarkonium spin observables as sensitive probes of the vortical structure and microscopic dynamics of the quark--gluon plasma.\\

\noindent
{\it Future Outlook:} Various directions can further advance the present study:

\begin{itemize}
\item The present study employs a static fireball characterized by a constant average temperature. Incorporating a realistic 3+1D hydrodynamic evolution of the medium would provide a more realistic description of the QGP and further constrain the theoretical predictions.

\item In particular, the most sensitive model parameter of this analysis is the spin-relaxation time $\tau_s$. A more precise determination of $\tau_s$ requires incorporating the heavy-quark momentum evolution with the inclusion of collisional and radiative energy loss in the numerical simulation. Furthermore, for a vortical medium, the spin relaxation time of heavy quarks needs to be calculated from a field-theoretical perspective in a rotating system.

\item Furthermore, it would be worthwhile to establish an Einstein–Stokes-type relation connecting the spin diffusion coefficient with the dissipative parameters governing spin transport in relativistic hydrodynamics.

\end{itemize}
\vskip2ex


\section*{Acknowledgement}
We thank Sourav Dey, Amaresh Jaiswal and Raghunath Sahoo for helpful discussion. Bhagyarathi Sahoo acknowledges the financial aid from CSIR, Government of India. Bhagyarathi Sahoo gratefully acknowledges funding from the DAE-DST, Government of India, under the mega-science project \enquote{Indian Participation in the ALICE experiment at CERN}, bearing Project No. SR/MF/PS-02/2021-IITI (E-37123). Captain R. Singh acknowledges the Department of Atomic Energy (DAE), India, for financial support. 


\vspace{10.005em}

\appendix

\section{Ensemble averaged $\mathbf{\omega}$ and $\mathbf{\rm B}$ field}
\label{appexA}
The primary source of vorticity in relativistic heavy ion collisions is the initial orbital angular momentum. The dominant contribution to the magnetic field comes from the charged spectators. Both the vorticity and magnetic field are predominantly directed along the $y$-direction. The electric field $\boldsymbol{E_{\rm Lab}}$ and electric-like component for rotation $\boldsymbol{k_{\rm Lab}}$ are assumed to be along the $x$-direction in the lab frame. If the initially produced heavy quarks are assumed to have an isotropic velocity distribution, then, upon averaging over the heavy-quark distribution, one can express

\begin{align}
\langle \boldsymbol{\omega} \rangle
&= \int \frac{d^{2}\Omega_v}{4\pi}\,
f_Q(\theta_v,\phi_v)
\Bigg[
\gamma_v
\left(
\boldsymbol{\omega}_{\rm Lab}
-
\boldsymbol{k}_{\rm Lab}\times\boldsymbol{v}
\right)
\nonumber\\
&\qquad\qquad
+ (1-\gamma_v)
\left(
\frac{\boldsymbol{\omega}_{\rm Lab}\cdot\boldsymbol{v}}
{|\boldsymbol{v}|^2}
\right)\boldsymbol{v}
\Bigg],
\label{appe1}
\end{align}

\begin{align}
\langle \boldsymbol{B} \rangle
&= \int \frac{d^{2}\Omega_v}{4\pi}\,
f_Q(\theta_v,\phi_v)
\Bigg[
\gamma_v
\left(
\boldsymbol{B}_{\rm Lab}
-
\boldsymbol{E}_{\rm Lab}\times\boldsymbol{v}
\right)
\nonumber\\
&\qquad\qquad
+ (1-\gamma_v)
\left(
\frac{\boldsymbol{B}_{\rm Lab}\cdot\boldsymbol{v}}
{|\boldsymbol{v}|^2}
\right)\boldsymbol{v}
\Bigg].
\label{appe2}
\end{align}
where $\theta_v$ and $\phi_v$ are the angles denoting the direction of the initial heavy quark velocity with respect to the fixed axis along $\boldsymbol{\omega_{\rm Lab}}$ and $\boldsymbol{B_{\rm Lab}}$.  $d^{2}\Omega_v = \sin\theta_v d\theta_v d\phi_v$ is the solid angle element. Assuming a uniform angular distribution of the initial heavy quark velocity, i.e., $f_{Q} (\theta_v, \phi_v)$ = const., the average over the vorticity and magnetic field can be expressed as;
\begin{equation}
    \langle \boldsymbol{\omega} \rangle = \frac{1}{3} (1+2\gamma_v) \boldsymbol{\omega_{\rm Lab}},
    \label{appe3}
\end{equation}
\begin{equation}
    \langle \boldsymbol{B} \rangle = \frac{1}{3} (1+2\gamma_v) \boldsymbol{B_{\rm Lab}}.
    \label{appe4}
\end{equation}
Therefore, the ensemble average over all possible heavy-quark polarization reveals that the resulting polarization is independent of the electric-like components of both the spin-polarization tensor and the field-strength tensor. In contrast, the magnetic field and vorticity fields leave a clear imprint on the polarization in the laboratory frame. The enhancement of heavy quark and hadron polarization at high-$p_{\rm T}$ can be understood from the functional dependence given in Eqs.~\ref{appe3} and ~\ref{appe4}. In particular, the Lorentz factor $\gamma_{v} = 1/\sqrt{1-\boldsymbol{v^{2}}} \equiv E/m_Q = (m_{\rm T}\cosh y)/m_Q$, exhibits a strong dependence on particle kinematics such as hadron momentum and rapidity. Consequently, the investigation of quarkonium polarization within this formalism becomes increasingly challenging at large forward rapidities, where Lorentz-boost effects are significantly enhanced.

\end{document}